\documentclass[reprint,aps,numerical,superscriptaddress,floatfix,nofootinbib]{revtex4-1}
\usepackage{amssymb}
\usepackage{amsthm}
\usepackage[figuresright]{rotating}
\usepackage{graphicx}
\usepackage{tikz}			
\usetikzlibrary{decorations.pathmorphing}       
\usetikzlibrary{decorations.pathreplacing}      
\usetikzlibrary{positioning}                    
\usetikzlibrary{arrows}                         
\usetikzlibrary{patterns}                       
\usetikzlibrary{shapes}                         
\usetikzlibrary{scopes}
\usepackage{textcomp}
\usepackage[T1]{fontenc}
\usepackage[tight]{units}

\newcommand{\figref}[1]{figure~\ref{#1}}
\newcommand{\tabref}[1]{table~\ref{#1}}

\usepackage{hyperref}
\begin{document}
\title{A CsI low temperature detector for dark matter search}
\newcommand{\LNGS}{\affiliation{INFN - Laboratori Nazionali del Gran Sasso, Assergi (AQ) I-67010 - Italy}}
\newcommand{\mpi}{\affiliation{Max-Planck-Institut f\"ur Physik, D-80805 M\"unchen - Germany}}
\newcommand{\AQ}{\affiliation{Dipartimento di Scienze Fisiche e Chimiche - Universit\`{a} degli studi dell'Aquila, I-67100 Coppito (AQ) - Italy}}
\newcommand{\GSSI}{\affiliation{Gran Sasso Science Institute, I-67100 L'Aquila - Italy}}
\newcommand{\Ukraine}{\affiliation{Institute for Scintillation Materials, U-61001 Kharkov - Ukraine}}
\newcommand{\INFNMilano}{\affiliation{INFN - Sezione di Milano Bicocca, Milano I-20126 - Italy}}
\newcommand{\Bicocca}{\affiliation{Dipartimento di Fisica, Universit\`{a} di Milano-Bicocca, Milano I-20126 - Italy}}
\newcommand{\INFNRome}{\affiliation{INFN - Sezione di Roma I, I-00185 Roma - Italy}}

\newcommand{\Vienna}{\affiliation{Institut f\"ur Hochenergiephysik der \"Osterreichischen Akademie der Wissenschaften, A-1050 Wien - Austria and Atominstitut, Vienna University of Technology, A-1020 Wien - Austria}}
\author{G.~Angloher}
  \mpi

\author{I.~Dafinei}
\INFNRome

\author{A.~Gektin}
\Ukraine

\author{L.~Gironi}
\INFNMilano
\Bicocca

\author{C.~Gotti}
\INFNMilano

\author{A.~G\"utlein}
\Vienna

\author{D.~Hauff}
 \mpi

\author{M.~Maino}
\INFNMilano

\author{S.S.~Nagorny}
 \GSSI

\author{S.~Nisi}
\LNGS

\author{L.~Pagnanini}
\GSSI

\author{L.~Pattavina}
 \LNGS

\author{G.~Pessina}
\INFNMilano

\author{F.~Petricca}
 \mpi

\author{S.~Pirro}
 \LNGS

\author{F.~Pr\"obst}
 \mpi
 
\author{F.~Reindl}
\email[corresponding author:]{florian.reindl@mpp.mpg.de}
\mpi

\author{K.~Sch\"affner}
\email[corresponding author:]{karoline.schaeffner@lngs.infn.it}
\GSSI

 \author{J.~Schieck}
 \Vienna
 
\author{W.~Seidel}
 \mpi
 
\author{S.~Vasyukov}
\Ukraine

\begin{abstract}
Cryogenic detectors have a long history of success in the field of rare event searches. In particular scintillating calorimeters are very suitable detectors for this task since they provide particle discrimination: the simultaneous detection of the thermal and the light signal produced by a particle interaction in scintillating crystals allows to identify the nature of particle, as the light yield depends thereon. We investigate the performance of two large CsI (undoped) crystals (\unit[$\sim$122]{g} each) operated as scintillating calorimeters at milli-Kelvin temperatures in terms of calorimetric properties and background rejection capabilities. Furthermore, we discuss the feasibility of this detection approach towards a background-free future dark matter experiment based on alkali halides crystals, with active particle discrimination via the two-channel detection.
\end{abstract}
\maketitle
\section{Introduction}
\label{intro}
Using low temperature detectors for the study of nuclear phenomena was first proposed by F.~E.~Simon, 80 years ago \cite{Simon}. Since then, considerable effort was put into the development of low temperature calorimeters for the search of rare events.\\
Present-date, the quest of the nature of dark matter is an open question of pivotal importance in the field of astroparticle physics. WIMPs (Weakly Interacting Massive Particles) with a mass in the (GeV-TeV)/c$^2$ regime and weak-scale interactions are a favorite class of possible dark matter candidates since they provide for the correct relic density \cite{goodman_detectability_1985}. Weak interaction allows for direct detection via elastic scattering on atomic nuclei (X,SM) $\rightarrow$ (X,SM) \cite{Bertone}. Since the expected recoil energies are very small, highly sensitive detectors are needed as well as an ultra-low background condition due to the expected small event rates. \\
The CRESST collaboration employs cryogenic scintillating calorimeters based on CaWO$_{\text{4}}$ \cite{CRESST1, CRESST2} which recently did prove an energy threshold for nuclear recoils as low as \unit[307]{eV} \cite{CRESST_LISE} combined with an excellent energy resolution ($\sigma_{0}$=\unit[62]{eV}) and particle discrimination. Such detectors are most suitable for the investigation of possible dark matter candidates, in particular covering also sensitivity for light dark matter particles thanks to their low energy threshold, a unique feature of the calorimetric technique.\\
We investigate the performance of a cryogenic calorimeter based on CsI (undoped), a crystal belonging to the family of alkali halides scintillators, regarding calorimetric properties (energy threshold and energy resolution) as well as particle discrimination via the simultaneous detection of scintillation light by employing a cryogenic light detector.\par
In the dark matter community alkali halide-based detectors (DAMA/LIBRA \cite{DAMA2013} ANAIS \cite{ANAIS}, DM-Ice \cite{DMICE2014}, KIMS \cite{KimIBS_NaI, KIMS2012}, SABRE \cite{SABRE2015}, PICO-lon \cite{PICOLON}) are so far only operated as single-channel devices (scintillation light only), thus not exploiting an active particle identification technique via an additional and independent channel.\\
We believe a scintillating calorimeter based on alkali halide material which provides a particle identification on an event-by-event basis by simultaneously detecting the thermal and the light signal produced by an interacting particle, is a powerful tool to study material-dependent interactions and allows to suppress background. These are two distinct requirements of future dark matter detectors utilizing alkali halides as target.
\section{Experimental Set-up}
\label{sec:2}
\begin{table}
\caption{Properties of undoped CsI and CaWO$_{\text{4}}$ \cite{Woody,Boyle,Nadeau,Schotanus,Zdesenko,Senyshyn}.}
 \begin{tabular}{lcccc}\hline
 \textbf{Properties} &  \textbf{CsI} & \textbf{CaWO$_{\text{4}}$} \\ 
 \hline
 Density [g/cm$^{\text{3}}$] & 3.67 & 6.12\\
 Melting point [$^{\circ}$C] & 661  & 1650\\ 
 Structure  & Cubic & Scheelite \\
 $\lambda_{\text{max}}$ at \unit[300]{K} [nm] & $\sim$300   & 420-425\\
 Hygroscopic & slightly & no\\
 $\Theta_{\text{D}}$ [K] & 169  & 335 \\
 Hardness [Mohs] &  2 & 4.4-5\\
 Mean energy of emitted photon [eV] & 3.3  & 3.1\\
 \hline
 \end{tabular}
\label{tab:crystal_properties}
\end{table}

\begin{figure}
\centering
  \begin{tikzpicture}[scale=0.8, every node/.style={scale=0.85}]
   
    \draw[color=black, thick, fill=black, opacity=0.13]
      (-1.5,-2.85) rectangle (3.5,1.85);
    \draw [color=black, thick, opacity=0.7]
      (0.5,2.7) -- (1.5,2.7);
    \draw[decorate, decoration={zigzag, pre length=5, post length=10, segment length=5}]
      (1,2.7) -- (1,1.6);
    \draw[color=blue, fill=blue, opacity=0.4]
      (-1.0,1.5) rectangle (3.0,1.35);
    \draw [color=black, pattern=north east lines]
    (0.5,1.6) rectangle (1.5,1.5);
    \draw[color=red, fill=red, opacity=0.65]
      (-.60,0.9) rectangle (2.6,-2.1);
    \draw[color=black, fill=black, opacity=0.2]
    (0.75,-2.1) rectangle (1.25,-2.2);
    \draw[color=orange, fill=orange, opacity=0.65]
      (-1,-2.35) rectangle (3,-2.21);
    \draw[decorate, decoration={zigzag, pre length=10, post length=5, segment length=5}]
      (1,-2.45) -- (1,-3.65);
    \draw[color=black, thick, opacity=0.7]
      (0.5,-3.65) -- (1.5,-3.65);
 \draw[color=black, pattern=north west lines]
      (3.25,-1.1) rectangle (3.45,-0.6);      
      
    \draw[color=black, pattern=north east lines]
    (0.5,-2.35) rectangle (1.5,-2.45);
    \draw[color=black,->] (4.0,2.7) node[color=black, right]{Thermal link} -- (1.7, 2.7);
    \draw[color=black,->] (4.0,2.1) node[color=black, right]{TES - light detector} -- (1.6, 1.7);
    \draw[color=black,->] (4.0,1.4) node[color=black, right]{Light absorber} -- (3.1, 1.4);
    \draw[color=black,->] (4.0,0.5) node[color=black, right]{CsI(undoped)} -- (2.3, 0.5);
    \draw[color=black,->] (4.0,-0.1) node[color=black, right]{Reflective foil} -- (3.6,-.1);
      \draw[color=black,->] (4.0,-0.85) node[color=black, right]{$^{224}$Ra/$^{55}$Fe source} -- (3.6,-0.85);
    \draw[color=black,->](4.0,-3.0) node[color=black, text width=2.3cm, right]{TES - crystal} --(1.5, -2.55);    
    \draw[color=black,->](4.0,-1.65) node[color=black, text width=2.3cm, right]{Silicon oil} --(1.22, -2.13);
    \draw[color=black,->](4.0,-2.3) node[color=black, text width=2.5cm, right]{Carrier crystal} --(3.05, -2.3);
    \draw[color=black,->](4.0,-3.65) node[color=black, text width=2.3cm, right]{Thermal link} -- (1.7, -3.65);
    \draw(-2.1,-0.6) node[color=red,thick, opacity=0.7, text width=2.5cm]{\textbf{Phonon}};
    \draw(-2.1,-1.0) node[color=red, thick, opacity=0.7, text width=2.5cm]{\textbf{detector}};
    \draw(-2.1,1.5) node[color=blue, thick, opacity=0.6, text width=2.5cm]{\textbf{Light}};
    \draw(-2.1,1.1) node[color=blue, thick, opacity=0.6, text width=2.5cm]{\textbf{detector}};
  \end{tikzpicture}
  \caption{Schematics of the detector module consisting of an undoped CsI target crystal and a light detector. Both detectors are read out by transition edge sensors (TES) and are surrounded by a reflective foil.}
\label{pic:module_schema}
\end{figure}
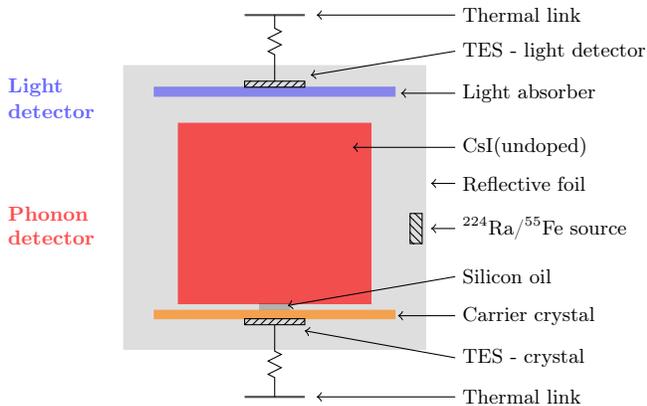
In this work we present the results from measurements of two different commercial \textit{undoped} CsI crystals operated as scintillating calorimeters at milli-Kelvin temperatures: CsI-Hilger and CsI-ISMA both of cubic shape with a side length of \unit[30]{mm} and a mass of about \unit[122]{g} each. All flat surfaces of both crystals were polished to optical quality.\footnote{A more precise discussion on the surface treatment is given later in section \ref{Ra_results}.} The crystal named CsI-Hilger was acquired from Hilger crystal company (UK), the CsI-ISMA was produced at Institute for Scintillating Materials in Kharkov, Ukraine. In \tabref{tab:crystal_properties} we list important properties of undoped CsI and of the crystal scintillator calcium tungstate (CaWO$_{4}$, used in CRESST-II) for comparison.\\
CsI is a very "soft" crystal.  Any mechanical contact with the crystal may induce stress to the crystal lattice, creating relaxation events that can mix up with particle events in the CsI operated as calorimeter. Thus, we designed a contact-free holding structure where the CsI crystal is attached to a  so-called \textit{carrier crystal}, a thin disc (\unit[40]{mm} in diameter, \unit[1.5]{mm} in thickness) made from CdWO$_{4}$. The contact between CsI and carrier crystal is realized by means of silicon oil. The mass ratio CsI to carrier crystal is 8.1.\\
The carrier crystal is kept in its copper structure via three bronze clamps and is equipped with a highly sensitive thermometer: a transition edge sensor (TES) consisting of a thin tungsten film (\unit[200]{nm}, W-TES) directly evaporated onto the carrier disc. Particle interactions in the CsI create tiny temperature excursions $\mathcal{O}$($\mu$K) which are measured by the change in the resistance of the TES. A dedicated heater stabilizes the W-TES in its transition between the normal and the superconducting phase. The heater consists of a gold stitch bond (gold wire with diameter of \unit[25]{$\mu$m}), which is bonded to a tiny gold structure on the TES which is at the same time used as thermal link.\par
The CsI crystal is facing a cryogenic light detector of CRESST-II type~\cite{CRESST1}, which consists of a thin sapphire disc (thickness of \unit[460]{$\mu$m} and diameter of \unit[40]{mm}). A \unit[1]{$\mu$m} thick layer of silicon is epitaxially grown onto the sapphire disc to absorb the blue scintillation light of CsI. Light detectors of this type we refer to as \textit{SOS light detector} (Silicon on Sapphire).\\
Just like the CsI, also the light absorber is read out by a W-TES, which in shape and dimension is optimized for the light detector. Crystal and light detector are enclosed in a reflective housing (Lumirror$^{\circledR}$) to maximize the light collection efficiency. The dedicated W-TESs were produced in collaboration with the Max-Planck-Institute for Physics in Munich, Germany. A detailed scheme of the \textit{detector module}, the ensemble of cryogenic light detector and scintillating CsI crystal, is depicted in \figref{pic:module_schema}.\par
The whole series of measurements was carried out in the test facility of the Max-Planck-Institute for Physics at the Laboratori Nazionali del Gran Sasso (LNGS), a deep underground site (\unit[3600]{m.w.e.}\cite{LNGS_muon}) in central Italy. The facility is composed of a dilution refrigerator which is  laterally surrounded by about \unit[20]{cm} of low-background lead to reduce the environmental $\beta$/$\gamma$-radioactivity. In order to ensure a low noise condition a platform, attached by a spring to the mixing chamber of the dilution unit, mechanically decouples the detector from the cryogenic facility. The set-up is a double stage pendulum, similar to the one described in \cite{Pirro}. The TESs are operated with a commercial dc-SQUID electronics (Applied Physics Systems). The hardware-triggered signals are sampled in a \unit[164]{ms} window and a sampling rate of \unit[50]{kHz}. Both detectors are always read out simultaneously, no matter of which one triggered.\par
For a detailed descriptions of the DAQ, the control of detector stability and the pulse height evaluation and energy calibration procedures the reader is referred to \cite{CRESST09,CRESST05}.

\section{Detector Performance}
\label{sec:3}
\subsection{Data processing}
Both crystals are mounted according to the set-up shown in \figref{pic:module_schema}. The crystals in all measurements are permanently irradiated from the lateral side with an $^{55}$Fe X-ray source. Additionally, an external and removable $^{241}$Am-source emitting \unit[59]{keV} $\gamma$-rays is temporarily used for a $\gamma$-calibration.\\
In one particular  measurement dedicated to study particle discrimination, a Ra-source is additionally placed close to the crystal, in position identical to the $^{55}$Fe-source.\\
Heater pulses are injected at regular time intervals of \unit[4]{s} to control stability, to linearize the detector response and to calibrate the detectors down to threshold energy. The overall count rate of these detector modules was about \unit[1]{Hz}, dominated by natural radioactivity arising from the non-radiopure materials of the cryostat.\\
Three basic cuts are applied to our data: right-left baseline parameter, pulse shape differences with respect to the template pulse via the RMS of the template fit, and rise time of the pulse via the peak position-onset parameter. The right-left baseline parameter is defined as the difference between the average level of the last and of the first 50 samples of a record. It allows to remove a large part of so-called decaying baselines which follow pulses induced by large energy depositions which do not relax back to equilibrium until the reactivation of the trigger. Additionally, this parameter discards artifacts from the readout electronics causing a change of the baseline level within the record. A cut on the RMS of the template fit of the registered particle pulse removes any pulse deviating from the nominal pulse shape (e.g. events with a strongly tilted baseline) and/or an enhanced noise level. Moreover, the class of events that certainly has to be identified and removed from the data are events originating from interactions taking place in the carrier,  so-called \textit{carrier events}. The fraction of carrier events that survives the RMS cut can be alternatively spotted by the rise time of the pulse, via the peak position-onset parameter.
\subsection{Calibration}
\label{LD}
The CsI crystal is calibrated with $\gamma$s from an externally applied $^{241}$Am-source. In the case of the light channel an equivalent energy of \unit[59.5]{keV$_{ee}$} is assigned to the scintillation peak in the energy spectrum of the light detector induced by the $^{241}$Am $\gamma$s in the CsI. To denote that the scale is only valid for electron recoils producing scintillation light in the CsI the subscript ee (electron-equivalent) is added to the energy scale.\footnote{For other particle types the respective QFs have to be considered.}\\
To additionally allow for a direct energy calibration of the light detector a weak $^{55}$Fe X-ray source (EC with subsequent de-excitation via X-ray emission) was faced  directly to the light absorber. The energy resolution achieved for the $^{55}$Mn K$_{\alpha}$- and K$_{\beta}$-lines is about $\sigma$=\unit[90.7]{eV}. The RMS of the baseline of the light detector is about $\sigma$=\unit[20]{eV}.\footnote{Such kind of light detectors used in CRESST-II experiment have demonstrated an RMS of baseline as good as $\sigma\sim$\unit[5]{eV}. Thus, it is possible to further improve the performance of the light detector by improving the electronic noise condition at the test facility.} The same light detector was utilized for all data presented within this work.

\subsection{Results of the CsI-Hilger Crystal}
\label{Hilger}
\begin{figure}
  \includegraphics[width=0.47\textwidth]{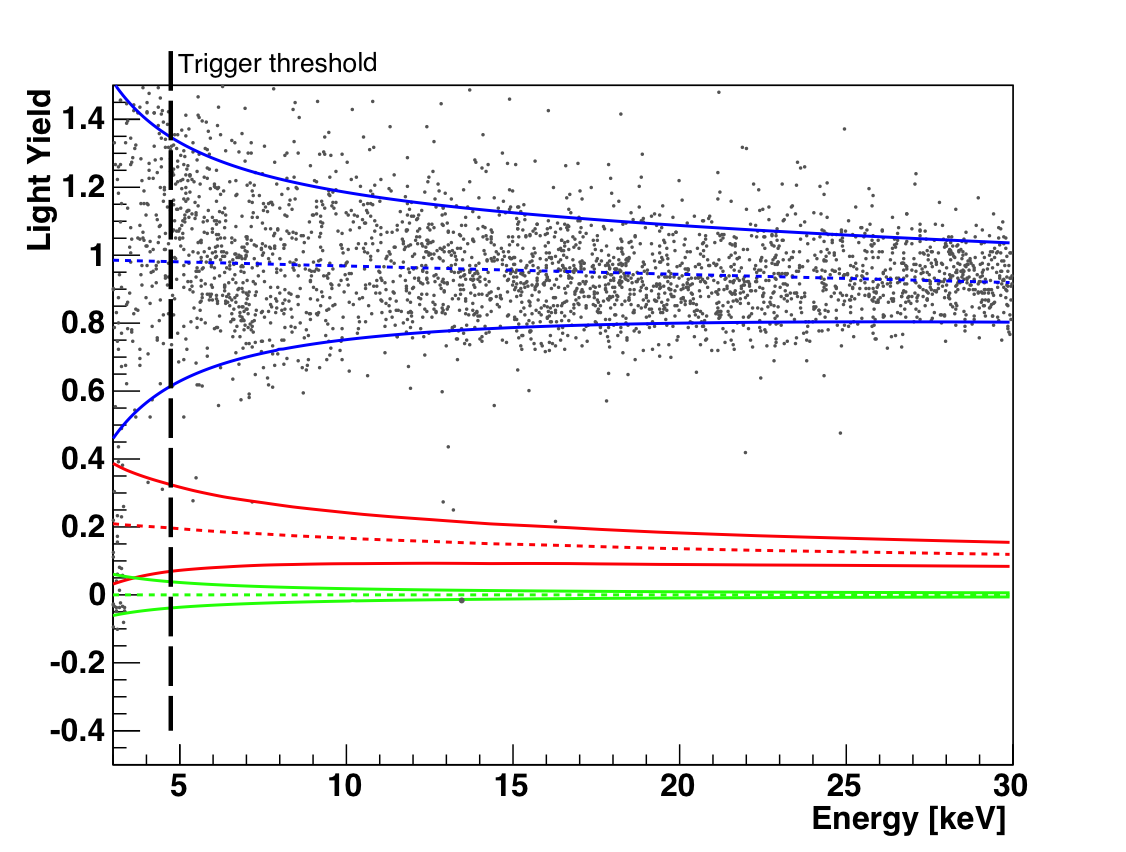}
 \caption{Background data of an undoped CsI crystal (0.458 kg-days of exposure; CsI-Hilger) in the light yield-energy plane. In between the solid lines 80\% of the events of the corresponding event class are expected; blue for electron recoils, red for nuclear recoils off Cs or I, green for potential events not producing scintillation light. The dashed lines mark the mean values of the central 80\% probability bands. A trigger threshold  as low as \unit[4.7]{keV} is achieved within the present detector set-up, see text for detailed information.}
\label{fig:LY_CsI}    
\end{figure}
\begin{figure}
  \includegraphics[width=0.47\textwidth]{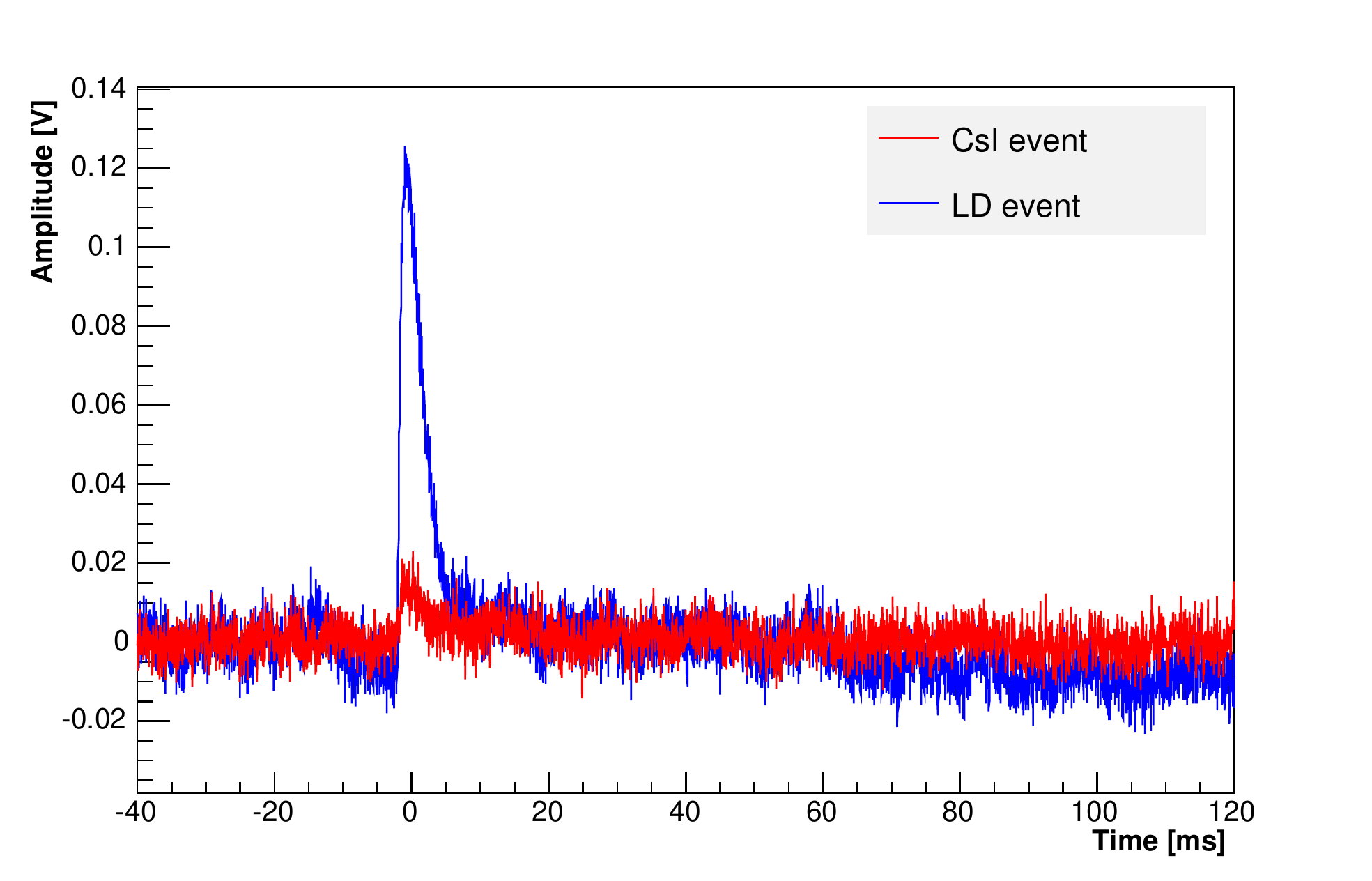}
  \caption{Coincident pulse as recorded with the CsI detector module. The phonon and light signal are colored in red and blue respectively. This event is due to an X-ray from an $^{55}$Fe source and being absorbed in the CsI crystal.}
\label{fig:55Fe_event}    
\end{figure}

Figure \ref{fig:LY_CsI} depicts about \unit[0.458]{kg-days} of background data in the light yield versus energy plane. Light yield we define here as the ratio of energy detected in the light detector expressed in keV$_{\text{ee}}$ (electron-equivalent) and the energy deposited in the CsI crystal in keV.\footnote{Since the energy measured in the crystal is for this measurement practically independent of the particle type the index \textit{ee} can be dropped for the phonon detector.} Thus, electron recoils get assigned a light yield of one. For other particle types, as mentioned later in the text, the respective quenching factors, which quantify the reduction in light output, have to be considered.\\
\begin{figure}
  \includegraphics[width=0.47\textwidth]{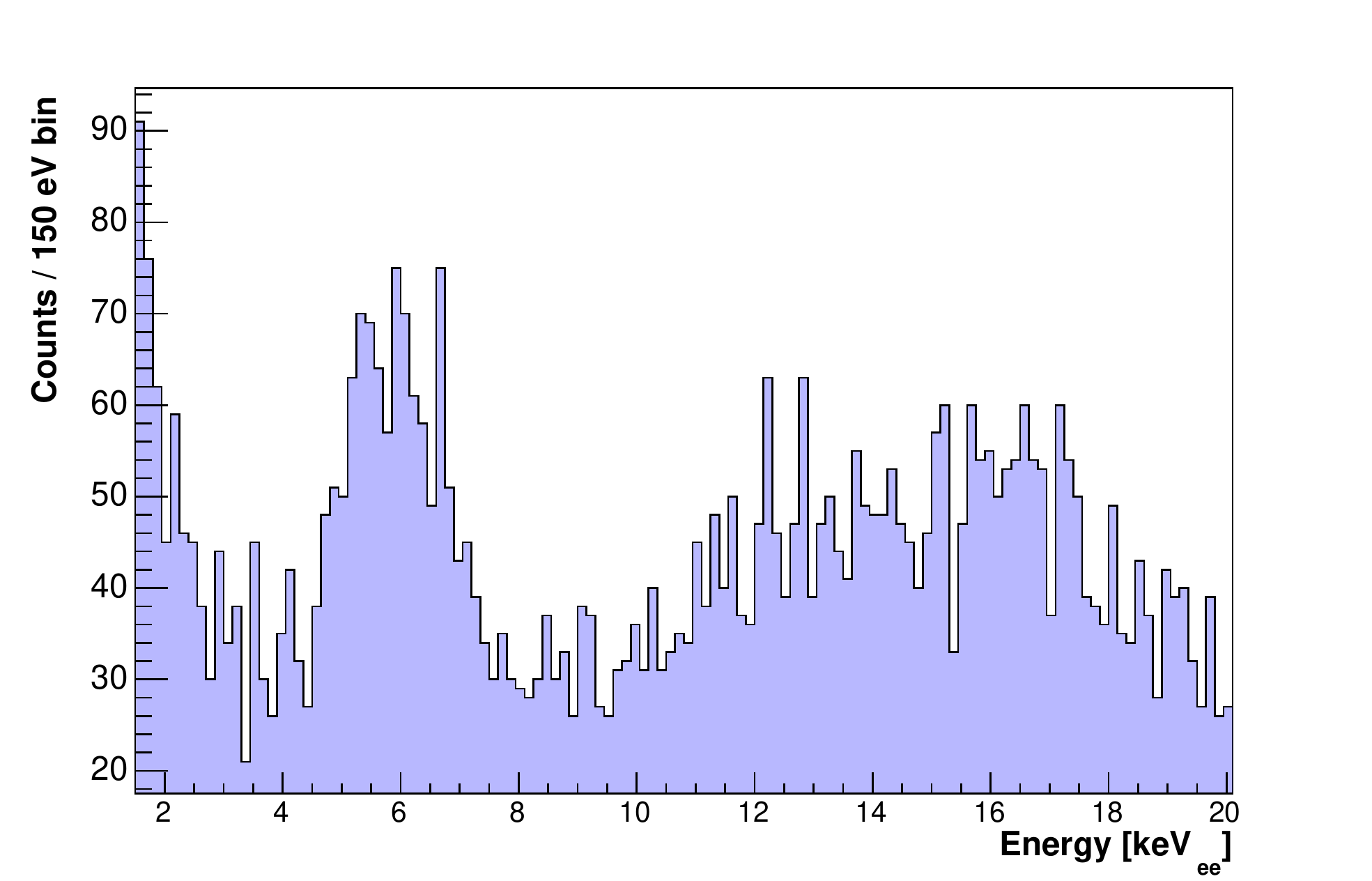}
  \caption{Energy spectrum of the SOS light detector for the scintillation light produced due to particle interactions in the CsI-Hilger crystal. The peak at around \unit[6]{keV$_{ee}}$ is due to the scintillation light emitted from the CsI-crystal due to the absorption of X-rays originating from a dedicated $^{55}$Fe X-ray source. The data correspond to an exposure of about \unit[0.986]{kg-days} (sum of background and calibration data).}
\label{fig:LD_spectrum}    
\end{figure}
The regions in the light yield-energy plane where different event-types are expected are described by bands. A band is defined by two functions, both depending on energy: the \textit{mean light yield} depending on the type of particle (quenching) and the \textit{width} set by the finite resolution of the detectors. Since the e/$\gamma$-band is highly populated, \textit{mean} and \textit{width} can be determined precisely by a fit to the data. With the parameters found for the e/$\gamma$-band and the quenching factor (QF) for the various event types the corresponding quenched bands are calculated. The QF is defined as the ratio of scintillation light produced by an interaction of a particle of a certain type to the scintillation light produced by a $\gamma$ of the same energy. We restrict ourselves to this qualitative description of the bands since a more detailed explanation and the validity of this approach was already proven, e.g.~in \cite{CRESST1, StraussQF}.\\
In \figref{fig:LY_CsI} we show the bands where possible event classes are expected for the CsI-Hilger crystal. In between the solid lines 80\% of all events of the corresponding event class are expected; in blue for electron recoils, in red for nuclear recoils off Cs or I and in green for potential events not producing scintillation light.\footnote{Since the QF for Cs and I is very similar, only one band is shown for reasons of clarity.} The energy-dependent QFs for Cs- and I-recoils are taken from Tretyak et al.~\cite{Tretyak}.\\
In \figref{fig:LD_spectrum} we show an energy spectrum of the scintillation light detected by the SOS light detector. The peak at around \unit[6]{keV$_{ee}}$ originates from scintillation light emitted by the crystal due to its irradiation with X-rays from an $^{55}$Fe source (K$_{\alpha}$=\unit[5.9]{keV} and K$_{\beta}$=\unit[6.5]{keV}). The physical position of the source is indicated in \figref{pic:module_schema}. The light energy resolution at this peak is \unit[(727.3$\pm$6.3)]{eV$_{ee}$}.\\
The presence of an additional $^{55}$Fe X-ray source, oriented towards the light absorber, gives us the possibility to gain a direct energy calibration of the light detector. With this information we can calculate the total amount of energy detected in the light detector of our set-up. We find that for CsI-Hilger about 8.1\% of the energy deposited in the CsI crystal is detected in form of scintillation light.\\
The last paragraph of this section is dedicated to the discussion of the achievable energy threshold of the CsI crystal. The physical properties of alkali halides crystals per se are not optimal for the application as cryogenic calorimeters:  the production, propagation and thermalization of phonons,  is best for materials with high Debye temperature, hence materials with generally high melting point and hardness.\\
In cryogenic scintillating calorimeters using a crystal scintillator with a relative high Debye temperature, thus suitable phonon transportation properties,  the energy threshold of the detector module is typically driven by the performance of the phonon detector.\footnote{In a CRESST-II CaWO$_{4}$ detector with similar layout typically $\sim$2\% of the deposited energy in the crystal is detected in form of scintillation light for $\beta$/$\gamma$-events. Thus, the light detector has to have an increased sensitivity by a factor of about 100 for $\beta$/$\gamma$ interactions; for other particle types the QF has to be considered in addition.} In other words, particle discrimination in such calorimeters at some point breaks down due to the limited sensitivity of the light detector or rather the amount of produced scintillation light for the given particle type at given energy deposit.\\
For the CsI detector module the situation is somehow reversed. The amount of produced scintillation light is increased by a factor of about two in comparison to e.g. a CaWO$_{4}$ crystal. However, at the same time we observe generally small pulse amplitudes in the phonon channel, most probably due to the unfavorable physical properties of CsI (also see \tabref{tab:crystal_properties}).\\
The pulses depicted in \figref{fig:55Fe_event} are recorded simultaneously and correspond to the heat (red) and scintillation light (blue) signal due to an energy deposition of an $^{55}$Fe X-ray in the CsI crystal. It is intuitively clear from \figref{fig:55Fe_event} that the light detector has still a definite signal whereas the heat signal induced by the \unit[5.9]{keV} X-ray in the crystal is already close to baseline noise. 
To estimate the trigger threshold we use pulses which are created by superimposing a signal template, scaled to the desired energy, and empty baselines. The signal template is acquired by averaging a large number of pulses of the same deposited energy (usually from a $\gamma$-calibration peak) and, thus, provides a noise-free description of the detector response to an energy deposition. Determining the energy resolution from these simulated pulses is a precise measure of the impact of the baseline noise, which also is the decisive factor for the achievable trigger threshold. For the CsI-Hilger crystal we find a resolution of $\sigma$=(\unit[946]{eV}$\pm$\unit[5]{eV}(stat.)). This method was validated by the CRESST experiment \cite{CRESST_LISE} proving long-term stability for a threshold setting of five times the baseline resolution, which in our measurement corresponds to \unit[4.7]{keV}.\\
Alternatively, the baseline resolution of a cryogenic calorimeter can also be extracted from the energy resolution of injected heater pulses. This approach is more conservative, for the measurement discussed here we find an energy resolution of $\sigma$=\unit[979]{eV} for the smallest injected test pulse with an equivalent energy of \unit[7.9]{keV}.\\
The estimated threshold value of \unit[4.7]{keV} is indicated as a vertical dashed line in \figref{fig:LY_CsI}. The events observed at energies below are mostly attributed to triggers of the light detector which simultaneously toggles the read-out of the phonon detector. However, it should be emphasized that for events with only the light detector triggering, nuclear recoil events are strongly disfavored with respect to e/$\gamma$-events, due to the quenched light signal of nuclear recoils. 
\subsection{Results of CsI-ISMA Crystal}
\begin{figure}
  \includegraphics[width=0.47\textwidth]{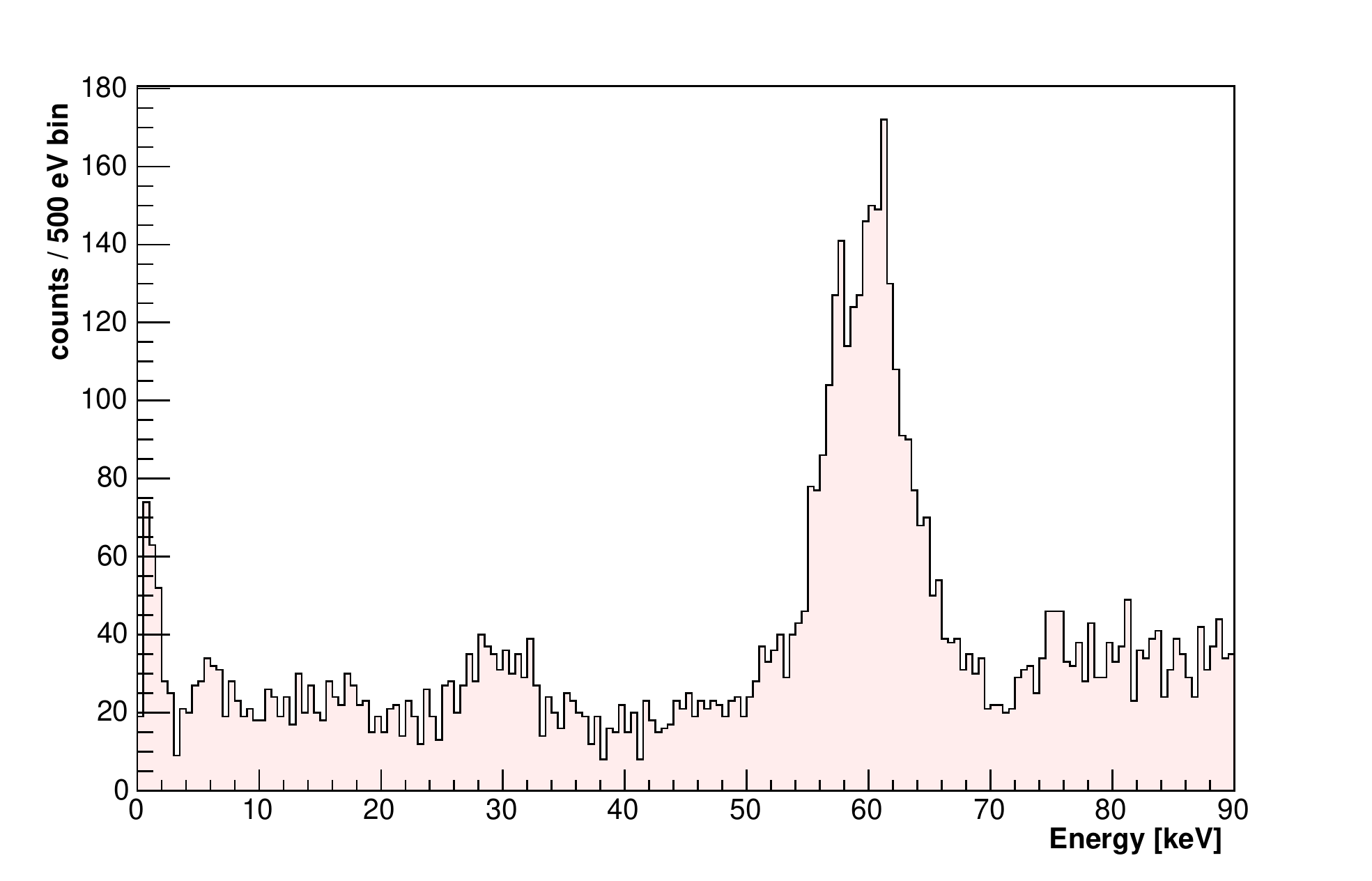}
  \caption{Energy spectrum of the deposited energy recorded in the CsI-ISMA crystal in about \unit[0.254]{kg-days} of exposure. The line from the external $^{241}$Am $\gamma$-source is detected at \unit[59.89]{keV} ($\sigma$=\unit[3.1]{keV}).}
\label{fig:CsI_spectrum}    
\end{figure}
The performance of the CsI-ISMA detector module is comparable to the CsI-Hilger detector discussed before. We interpret this as a positive sign, since the two crystals have been produced in different growth facilities, starting from different raw materials. Only the light detector was affected by some additional microphonic disturbances induced by the cryostat, resulting in a slightly worse energy resolution. The energy spectrum of the CsI-ISMA crystal acquired in about \unit[0.254]{kg-days} of exposure and in presence of an $^{241}$Am $\gamma$-source is shown in \figref{fig:CsI_spectrum}. A peak is observed with a fitted mean of  \unit[(59.89$\pm$0.10)]{keV} ($\sigma$=\unit[(3.11$\pm$0.14)]{keV}). This peak corresponds to the $^{241}$Am $\gamma$-line.\footnote{Literature value for gammas from $^{241}$Am is \unit[59.54]{keV}.}\par
We find the energy resolution of the smallest injected heater pulse, with an equivalent energy of \unit[7.0]{keV}, to be $\sigma$=\unit[702]{eV}. This value is slightly better than the resolution determined for the CsI-Hilger crystal. Consequently, the CsI-ISMA crystal exhibits a slightly lower estimated trigger threshold of about \unit[3.5]{keV}.
For CsI-ISMA the fraction of energy detected in form of scintillation light is about 6.5\%, thus $\sim$20\% less than the CsI-Hilger. Chemical impurities in the crystal are known to suppress scintillation in crystals \cite{Dafinei}. In section \ref{ICPMS} we discuss the presence of impurities in the CsI crystals on basis of the results from an ICP-MS analysis of samples taken from these crystals.\\
To summarize, the two cryogenic measurements carried out on CsI-Hilger and CsI-ISMA prove that CsI can be successfully operated as a scintillating cryogenic calorimeter with a threshold, in the case of e.g. CsI-Hilger as low as \unit[4.7]{keV} and for CsI-ISMA  as low as about \unit[3.5]{keV}, in the present set-up.
\section{Particle discrimination}
In order to study the response of the CsI to particles recoiling off the caesium and iodine target nuclei, the most suitable solution would be to irradiate it with neutrons from an external neutron source. Since, at the time of the measurement, such a source was not available, we decided for a first measurement with a source producing $\alpha$-particles and accordingly recoiling nuclei, the latter being of interest for the demonstration of particle discrimination in CsI. The crystal used for this measurement is CsI-Hilger. The experimental set-up is identical to the measurement described in section \ref{Hilger}, only a small aluminum foil (\unit[2]{cm$^{2}$}) carrying the source was inserted and attached onto the Lumirror$^{\circledR}$ reflector, as indicated in \figref{pic:module_schema}.\footnote{In comparison to VM2002 (3M company) the Lumirror$^{\circledR}$ material does not scintillate, but only works as a reflector.}
\subsection{Source}
The source was made by exposing an aluminum foil in vacuum conditions to a surface-implanted  $^{228}$Th-source, thus having a certain chance for implanting $^{224}$Ra-nuclei onto the aluminum foil (\cite{Alessandrello, Arnaboldi}). The penetration depth of a $^{224}$Ra-nucleus of e.g.~\unit[90]{keV} in aluminum is  $\mathcal{O}$(\unit[100]{nm}) \cite{Ziegler}. $^{224}$Ra, with a half life of \unit[3.66]{d} further $\alpha$-decays to $^{220}$Rn followed by $^{216}$Po, $^{212}$Pb, $^{212}$Bi and $^{212}$Po - the fifth and last $\alpha$-decay in the chain to stable $^{208}$Pb.\\
Given the sequence of $\alpha$-decays, there is a realistic probability to have nuclear recoils escaping from the aluminum foil. Especially $^{224}$Ra atoms which were implanted at a shallow depth have the highest probability to provide nuclei that can escape the source.\footnote{At the same time the source produces $\alpha$-particles with a higher rate and a good energy resolution of the $\alpha$-lines, thanks to the shallow implantation profile of the primary $^{224}$Ra-nuclei (max.$\sim$\unit[550]{nm}) being almost negligible in sense of energy loss considering few MeV $\alpha$s.} These slow and heavy recoils interacting in the CsI mimic the behavior of a recoiling nucleus of the lattice itself. The energy distribution of such recoil events is expected to reach from \unit[$\sim$100]{keV} down to threshold energy of the CsI detector. For a more precise answer a simulation would have to be performed on the various $\alpha$-decays and their density profile in the aluminum foil. However, without having information on the surface condition of the foil and the CsI crystal, thus not having in hand the correct starting parameters, we do not have a reliable basis in order to start such a complex simulation which completely relies on surface properties, due to the short range of nuclei in such material.
\begin{figure}
  \includegraphics[width=0.47\textwidth]{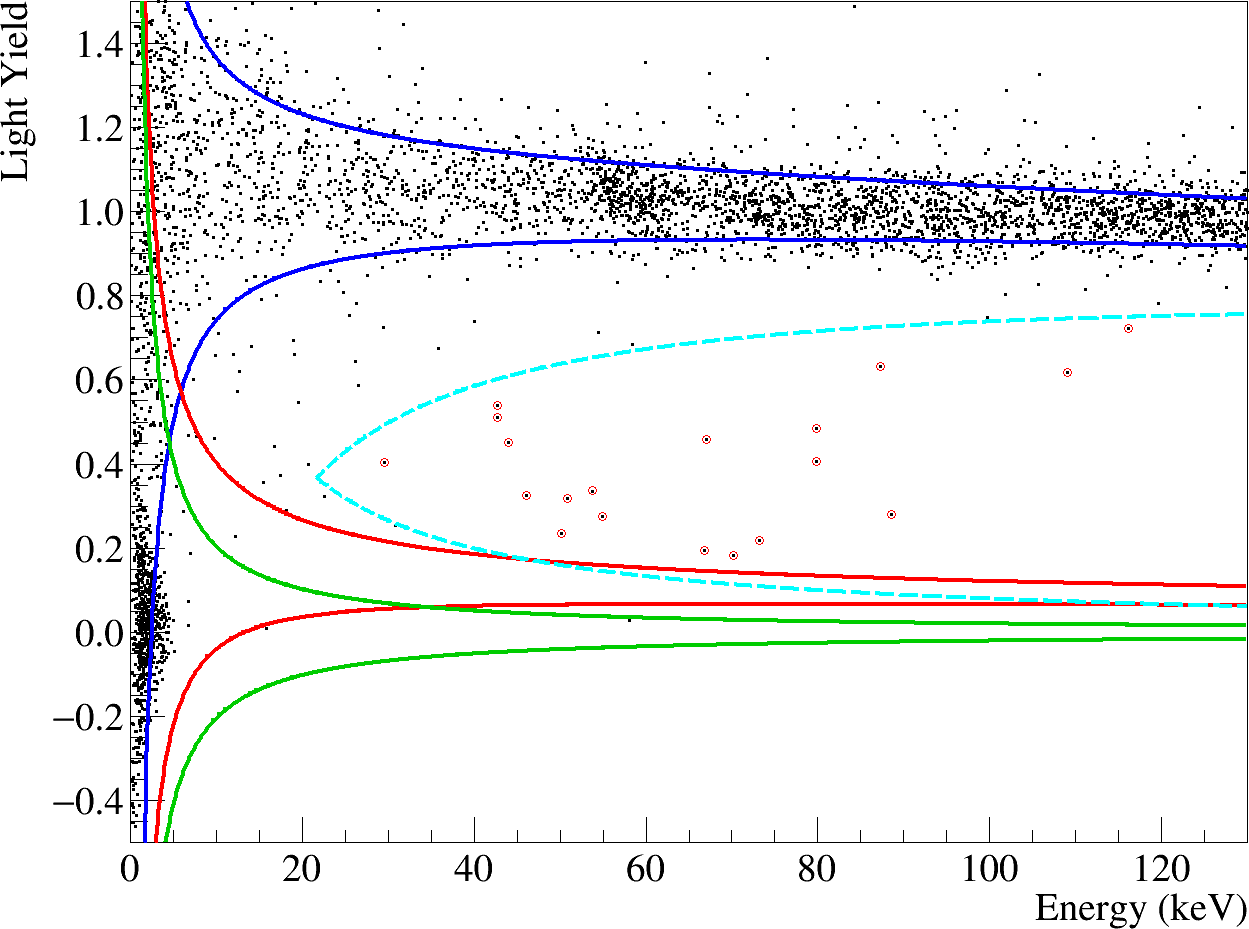}
  \caption{Data of CsI-Hilger in the light yield-energy plane of an exposure of about \unit[0.163]{kg-days}. During the measurement time the CsI was constantly exposed by $\gamma$s of an external $^{241}$Am source as well as by recoils induced by a $^{224}$Ra source, directly faced to the crystal (see \figref{pic:module_schema}). The bands are colored in analogy to figure \ref{fig:LY_CsI}, blue for e/$\gamma$-events, red for recoils off Cs and I and green for potential phonon-only events not producing scintillation light. The upper and lower cyan dashed lines mark the 5$\sigma$-boundaries of the e/$\gamma$-band and the phonon-only band. The events in between the cyan lines (marked with a red circle) are therefore statistically incompatible with leakage from the respective bands, thus providing reasonable evidence to be induced by heavy nuclei (Rn/Po from $^{224}$Ra source) interacting in the very near-surface layers of the CsI.}
\label{fig:PID}    
\end{figure}
\subsection{Experimental Results}
\label{Ra_results}
Combining both a theoretical approach to the detector response from nuclei as e.g.~Rn and Po and experience of QFs measured for such heavy nuclei in other scintillating materials at low temperatures \cite{CRESST1}, we would expect the recoil-events to accumulate in form of a narrow band at a low light yield value. Such a band for Cs and I recoils was already shown  in \figref{fig:LY_CsI}, whereby the energy-dependent QFs for Cs and I were adapted from Tretyak et al.~\cite{Tretyak}.\\
In \figref{fig:PID} data of an exposure of about \unit[0.163]{kg-days} of the CsI-Hilger module  is shown in the light yield-energy plane. Apart from the $^{224}$Ra-recoil source an external $^{241}$Am $\gamma$-source was present during the whole data taking for the energy calibration of the detectors.\footnote{Due to the short half life of $^{224}$Ra the whole measurement, including detector preparation, cool-down of the cryostat and the optimization of the detectors including energy calibration had to be carried out in a most efficient and least time consuming way. Furthermore, the operating points of the detectors were not optimized at the level of the previous measurement (\figref{fig:LY_CsI}), due to lack of time.}\\
Looking at \figref{fig:PID}, the highly populated event-distribution at a light yield (LY) around 1 can be assigned to e/$\gamma$-interactions in the CsI. The color code for the band description, blue for e/$\gamma$-events, red for recoils off Cs and I and green for potential phonon-only events not producing light is chosen identical to \figref{fig:LY_CsI}.  In addition, we display cyan-colored dashed lines which mark the lower and upper 5$\sigma$-boundary of the e/$\gamma$-band and the phonon-only band, respectively. Thus, the distribution of events (marked with a red circle) in between the cyan lines are statistically neither compatible with leakage from the e/$\gamma$-band, nor to the phonon-only events band.  Two characteristic features of these events may be observed. Firstly, the broad light yield distribution ($0.2\lesssim \text{LY} \lesssim 0.6$) is not in accordance with a band-like interpretation and, secondly, the energies of the major number of events does extend from threshold up to \unit[$\sim$90]{keV} only. Solely two events are present at slightly higher energies but significantly higher light yield.\\
Since this class of events (marked with a red circle)  is not present in the previous measurements without recoil-source (see \figref{fig:LY_CsI}, exposure of a factor 2.8 larger), detector specific artifacts and a strong leakage-component from e/$\gamma$-events can be excluded in advance, moreover as the same crystal was used. To evaluate if these events can be due to $^{220}$Rn recoiling off Cs or I nuclei, we may investigate the $\alpha$-particles produced by this source. In total we find 2106 $\alpha$-events in the complete data set from the four expected lines at \unit[5685]{keV}, \unit[6288]{keV}, \unit[6778]{keV} and \unit[8784]{keV}. The events in the cloud-like distribution count 19, thus a fraction of solely 0.9\%, in comparison to the total number of recorded $\alpha$-events. Recoiling nuclei can travel only $\mathcal{O}$(\unit[100]{nm}) in aluminum before they are stopped, however the range of $\alpha$-particles with the given energies is few $\mathcal{O}$(\unit[1]{$\mu$m}). Thus, given the range differences for nuclear recoils and $\alpha$-particles, we expect only such a small percentage.\\
That heavy nuclei interacting in the near-surface layers in undoped CsI can in fact reveal a peculiar LY is further supported by experimental observations discussed in \cite{Getkin92, Getkin95, Kudin}: they observe an intensive blue luminescence in CsI due to the formation of vacancies as a consequence of e.g.~plastic deformation of the crystal. The optically polished surface of our CsI sample implies a mechanical treatment of the surface, known to damage the near-surface layers of CsI. Such near-surface layers exceed - typically by many orders of magnitude - the equilibrium concentration of vacancies, which are known to change the correlation between the various components of the luminescence, resulting in an additional and intense luminescence contribution in the range of \unit[440-550]{nm} \cite{Getkin95}. Defect surface layers are difficult to remove, especially in case of absence of a cleavage plane, as in CsI. In \cite{Kudin} a surface-removing procedure is discussed which relies on \textit{two} important steps: first, after mechanical surface treatment of the CsI it has to be stored in ambient air (relative humidity above 30\%) to allow for recombination/relaxation of vacancies ($\mathcal{O}$(\unit[]{days})). Second, still remaining impurities and defects may be removed by chemical polishing using e.g.~Methanol.\\
The Hilger-CsI was undertaken a partially similar procedure: after mechanical polishing of the surface it was chemically treated with a mixture of alcohols in order to remove the defect layer induced by the machining. But, the relaxation time for near-surface vacancies in ambient air was not respected, hence, there is a certain chance to still have defect layers, even after the chemical polishing.\\
Thus, there is reasonable evidence that the events induced by our $^{224}$Ra recoil-source can create events in the CsI which are accompanied by a larger scintillation light signal, in particular since such nuclei are stopped in CsI within the first $\mathcal{O}$(100) atomic layers, depending on the initial energy of the nucleus. The short range of such nuclei is also an excellent measure for the overall homogeneity of the defect surface, present in our sample.\\
From the measurement we can also conduct an upper limit on the thickness of the defect layer present in the CsI samples. In all measurements we irradiate the CsI with X-rays from $^{55}$Fe. The absorption length of \unit[5.9]{keV} X-rays in CsI is $\sim$\unit[5]{$\mu$m}. As we do not observe a distorted scintillation peak from $^{55}$Mn K$_{\alpha}$ and K$_{\beta}$ in the light detector (see \figref{fig:LD_spectrum}) we can acknowledge that the defect near-surface layers does not exceed a thickness of about \unit[5]{$\mu$m} in CsI-Hilger and CsI-ISMA.\\
In the frame of introducing cryogenic calorimeters based on an alkali halide target for the search of dark matter particles such kind of surface effect is not of concern. First, as discussed previously, Kudin et al.~\cite{Kudin} started to develop a recipe to successfully remove defect surface layers. Second, $\alpha$-induced background has to be  prevented in any case, as nuclear recoils can mimic WIMP interactions. This can be achieved by using a completely scintillating housing \cite{TUM40} of the detector or an active shielding to tag the simultaneously emitted $\alpha$ in order to reject such events \cite{Becher}. Third, dark matter interactions are expected to take place in the volume, thus are not affected by very near-surface properties. From another point of view the observed increased surface luminescence may be used in order to reject any surface-related background, a powerful tool in rare event searches.\\
To summarize, we observe a distribution of events at low light yield values which in number are consistent with the expected number of recoils induced by the $^{224}$Ra recoil-source faced to the CsI. This is further supported by the observation of such events only up to an energy of about \unit[100]{keV}, the maximum energy a Ra-recoil can acquire from the decay. The broad distribution in light yield can have its explanation in the presence of defect layers which, as discussed in literature, are known to show an increased scintillation light production due to the high amount of available vacancies. Thus, there is room to interpret the observed event distribution being due to nuclear recoils, hence proving the capability of particle identification in CsI (undoped) by making use of the simultaneous detection of the thermal and the light signal.\\
However, a measurement with a neutron source is needed to quantify the discrimination power that can be achieved in this material. Such a measurement is indispensable in order to judge the potential of alkali halide crystal for rare event searches in detail.\\
At last we want to refer back to the background measurement of CsI-Hilger discussed in \ref{Hilger} and displayed in figure \ref{fig:LY_CsI}. In these data we find one event without an associated light signal at around \unit[14]{keV}, a so-called \textit{phonon-only event}. Since we expect an increased light yield from such surface nuclear recoil events  induced by surface $\alpha$-decays (see \ref{Ra_results} above), we expect this event to result from lattice relaxation due to the not perfect surface quality of the CsI-Hilger crystal.\\
There are few events detected slightly above the upper 80$\%$ line of the almost empty nuclear-recoil band (red) as well as three events around a light yield of 0.5. In fact, since no passive neutron shielding is available, we expect to see neutron-induced recoil events in our data set.\footnote{From a sequence of measurements carried out in the test cryostat the neutron count rate is estimated to 2-3 neutrons/(kg day).}
The overall high population of the e/$\gamma$-band has its explanation in the shielding situation of the test facility. The cryostat itself is not made from radiopure materials and the external \unit[20]{cm} lead shielding is not enclosing the set-up completely, but leaving the top part uncovered.
\section{Chemical purity of CsI crystals}
\label{ICPMS}
\begin{table}
\caption{Impurities in CsI-Hilger and CsI-ISMA from results of ICP-MS analyses. Uncertainties are at 30\%.}
\label{tab:ICPMS}   
\begin{tabular}{l l c c}
\hline\noalign{\smallskip}
Element & Unit &  CsI-Hilger & CsI-ISMA\\
\noalign{\smallskip} 
\hline
\noalign{\smallskip}
 Tl & [ppb]&    2.9     & 1.9   \\
 Na  & [ppm] &    0.16     & 0.92  \\
 \hline
  Ca  & [ppm] &    13.3     & 3.3 \\
   K  & [ppm] &    8.08   & 3.88 \\
   \hline
   Cr  & [ppb] &    <19     & <17  \\
    Fe  & [ppb] &    <375     & <167  \\
     Co  & [ppb] &    <2   & <2 \\
       Ni  & [ppb] &    <28   & <25 \\
        As  & [ppb] &    <8   & <7 \\
          Mn  & [ppb] &    <19  & <5 \\
\noalign{\smallskip}\hline
\end{tabular}
\end{table}
In order to investigate the amount of chemical contaminations in the CsI-Hilger and CsI-ISMA material we performed material screenings using Inductively Coupled Plasma Mass Spectrometry (ICP-MS) carried out on-site at LNGS.\\
Especially, we are interested in understanding to which extent the discrepancy in the observed amount of scintillation light for CsI-Hilger and CsI-ISMA ($\sim$20\%) originates from chemical impurities, known to harm the scintillation performance (mainly involving contributions from V, Cr, Mn, Fe, Co and Ni \cite{Dafinei}). Last but not least also the bolometric performance of the CsI can be affected by adding an additional contribution to its heat capacity (involving ferromagnetic and paramagnetic elements \cite{heatcap}).\\
Table \ref{tab:ICPMS} lists the measured concentrations of chemical impurities present in the two samples. We find very low limits, at ppb level, for Cr, Fe, Co, Ni, As and Mn. The amount of potassium is few ppm, as expected due to the chemical affinity of K and Cs.\footnote{Both crystals are commercial crystals without any specification on low radioactivity.}\\
In comparison to other elements, the content of Ca is different in the samples, with about 13 ppm in CsI-Hilger and about 3 ppm in CsI-ISMA. Doping of CsI with divalent ions (Me$^{2+}$), as e.g.~Ca, is known to introduce an additional blue emission band around \unit[415]{nm} \cite{CsI_Ca}, thus an enhancement in scintillation emission.\\
For what concerns the impurities from Tl and Na, typically used as dopants in CsI, we find only spurious traces of Tl in both samples. Thus, special care was taken in avoiding such contaminations by using for example a dedicated crucible. While doping CsI with Tl and Na is known to increase the light yield at room temperature, it will suppress the intrinsic scintillation, the mechanism responsible for the light production at low temperatures.\\
The measured amount of Na/Ca in CsI-ISMA is about a factor of $\sim$5 higher/ $\sim$10 lower than in CsI-Hilger. Though it should be mentioned that standard doping levels are in the range of 1000 ppm, thus three orders of magnitude higher. The presence of 1 ppm of Na and about 3 ppm of Ca can, from our understanding,  have only a marginal effect on the overall scintillation light production, possibly sufficient to explain the missing $\sim$20\% in CsI-ISMA in comparison to CsI-Hilger. 

\section{Perspective}
\label{sec:4}
In this manuscript we report the successful detector development and first results of a scintillating calorimeter using CsI as target, a crystal belonging to the family of alkali halides.\\
In the very near future we aim to switch from CsI to NaI, the target material used by the DAMA/LIBRA collaboration. DAMA/LIBRA observes a statistically robust annual modulation signal which can be interpreted as a signature for dark matter \citep{DAMA2013}. The development of a first prototype NaI scintillating calorimeter will be carried out within COSINUS (Cryogenic Observatory for SIgnatures seen in Next-generation Underground Searches), an R\&D project funded by CSN5 of Istituto Nazionale di Fisica Nucleare (INFN) and located at LNGS in Italy.\\
In case the performance of such NaI calorimeters can be proven to be comparable to already existing scintillating calorimeters (energy threshold of \unit[1]{keV} and lower), as e.g.~used in the CRESST dark matter search, such cryogenic NaI detectors have the potential to give an answer on the particle interaction channel participating in the DAMA/LIBRA modulation signal with higher sensitivity due to the significantly lower energy threshold for nuclear recoil signals and within a very moderate exposure of few \unit[10]{kg-days} \cite{COSINUS}.
\section*{Acknowledgments}
This work was supported by the Italian Ministry of Research under the PRIN 2010ZXAZK9 2010-2011 grant. We grateful acknowledge the generous support of LNGS of this activity.\\
In particular, we want to thank the LNGS mechanical workshop team E.~Tatananni, A.~Rotilio, A.~Corsi, and B.~Romualdi for continuous and constructive help in the overall set-up construction and M.~Guetti for his constant technical support in the underground facility.

\bibliographystyle{h-physrev}
\bibliography{CsI_arxiv.bib}

\begin{thebibliography}{10}

\bibitem{Simon}
F.~E. Simon,
\newblock Nature {\bf 135}, 763 (1935).

\bibitem{goodman_detectability_1985}
M.~W. Goodman and E.~Witten,
\newblock Physical Review D {\bf 31}, 3059 (1985).

\bibitem{Bertone}
G.~Bertone {\em et~al.},
\newblock {\em {Particle Dark Matter}} (Cambridge University Press, 2010).

\bibitem{CRESST1}
G.~Angloher {\em et~al.},
\newblock Eur. Phys. J. C {\bf 72}, 1971 (2012).

\bibitem{CRESST2}
CRESST Collaboration, G.~Angloher {\em et~al.},
\newblock {Eur. Phys. J. C} {\bf 74}, 3184 (2014), 1407.3146.

\bibitem{CRESST_LISE}
{CRESST} Collaboration, {Angloher, G.} {\em et~al.},
\newblock Eur. Phys. J. C {\bf 76}, 25 (2016).

\bibitem{DAMA2013}
{DAMA} Collaboration, R.~Bernabei {\em et~al.},
\newblock Eur. Phys. J. C {\bf 73} (2013).

\bibitem{ANAIS}
J.~Amar\'{e} {\em et~al.},
\newblock AIP Conference Proceedings {\bf 1672} (2015).

\bibitem{DMICE2014}
DM\char21{}Ice Collaboration, J.~Cherwinka {\em et~al.},
\newblock Phys. Rev. D {\bf 90}, 092005 (2014).

\bibitem{KimIBS_NaI}
K.~Kim {\em et~al.},
\newblock Astroparticle Physics {\bf 62}, 249  (2015).

\bibitem{KIMS2012}
KIMS Collaboration, S.~C. Kim {\em et~al.},
\newblock Phys. Rev. Lett. {\bf 108}, 181301 (2012).

\bibitem{SABRE2015}
E.~Shields, J.~Xu, and F.~Calaprice,
\newblock Physics Procedia {\bf 61}, 169  (2015),
\newblock 13th International Conference on Topics in Astroparticle and
  Underground Physics, TAUP 2013.

\bibitem{PICOLON}
K.~Fushimi {\em et~al.},
\newblock Journal of Physics: Conference Series {\bf 469}, 012011 (2013).

\bibitem{Woody}
C.~Woody {\em et~al.},
\newblock IEEE Trans. Nucl. Sci. {\bf 37}, 492 (1990).

\bibitem{Boyle}
A.~Boyle and G.~Perlow,
\newblock Phys. Rev. {\bf 151}, 211 (1966).

\bibitem{Nadeau}
P.~Nadeau {\em et~al.},
\newblock Astroparticle Physics {\bf 67}, 62  (2015).

\bibitem{Schotanus}
P.~Schotanus,
\newblock Nuclear Science, IEEE Transactions on {\bf 37}, 177 (1990).

\bibitem{Zdesenko}
Y.~Zdesenko {\em et~al.},
\newblock Nucl. Instr. Meth. Phys. Res. A {\bf 538}, 657 (2005).

\bibitem{Senyshyn}
A.~Senyshyn, H.~Kraus, V.~B. Mikhailik, and V.~Yakovyna,
\newblock Phys. Rev. B {\bf 70}, 214306 (2004).

\bibitem{LNGS_muon}
M.~Ambrosio {\em et~al.},
\newblock Phys. Rev. D {\bf 52}, 3793 (1995).

\bibitem{Pirro}
S.~Pirro,
\newblock Nucl. Instr. Meth. in Phys. Res. A {\bf 559}, 672  (2006).

\bibitem{CRESST09}
{CRESST} Collaboration, G.~Angloher {\em et~al.},
\newblock Astropart. Phys. {\bf 31}, 270 (2009), 0809.1829.

\bibitem{CRESST05}
{CRESST} Collaboration, G.~Angloher {\em et~al.},
\newblock Astropart. Phys. {\bf 23}, 325 (2005), astro-ph/0408006.

\bibitem{StraussQF}
{Strauss, R.} {\em et~al.},
\newblock Eur. Phys. J. C {\bf 74}, 2957 (2014).

\bibitem{Tretyak}
V.~Tretyak,
\newblock Astropart. Phys. {\bf 33}, 40 (2010).

\bibitem{Dafinei}
I.~Dafinei,
\newblock Journal of Crystal Growth {\bf 393}, 13  (2014).

\bibitem{Alessandrello}
A.~Alessandrello {\em et~al.},
\newblock Physics Letters B {\bf 408}, 465  (1997).

\bibitem{Arnaboldi}
C.~Arnaboldi {\em et~al.},
\newblock Astroparticle Physics {\bf 34}, 344  (2011).

\bibitem{Ziegler}
J.~Ziegler {\em et~al.},
\newblock {Stopping and Range of Ions in Matter},
\newblock \url{http://www.srim.org}, 2012,
\newblock SRIM-2012.03.

\bibitem{Getkin92}
A.~Gektin {\em et~al.},
\newblock {Opt. i Spektrosk.} {\bf 72}, 1061 (1992).

\bibitem{Getkin95}
A.~Gektin, I.~Krasovitskaya, N.~Shiran, V.~Shlyahturov, and E.~Vinograd,
\newblock Nuclear Science, IEEE Transactions on {\bf 42}, 285 (1995).

\bibitem{Kudin}
A.~M. Kudin, L.~A. Andryushchenko, V.~Y. Gres', A.~V. Didenko, and T.~A.
  Charkina,
\newblock J. Opt. Technol. {\bf 77}, 300 (2010).

\bibitem{TUM40}
R.~Strauss {\em et~al.},
\newblock The European Physical Journal C {\bf 75} (2015).

\bibitem{Becher}
F.~Reindl {\em et~al.},
\newblock Astroparticle, Particle, Space Physics and Detectors for Physics
  Applications - Proceedings of the 14th ICATPP Conference {\bf 8}, 290 (2014).

\bibitem{heatcap}
T.~O. Niinikoski, A.~Rijllart, A.~Alessandrello, E.~Fiorini, and A.~Giuliani,
\newblock EPL (Europhysics Letters) {\bf 1}, 499 (1986).

\bibitem{CsI_Ca}
S.~Myagkota, A.~Pushak, G.~Stryganyuk, S.~Novosad, and I.~Pashuk,
\newblock Functional Materials {\bf 15}, 187 (2008).

\bibitem{COSINUS}
G.~Angloher {\em et~al.},
\newblock in preparation  (2016).

\end{thebibliography}

\end{document}